\RequirePackage[2020-02-02]{latexrelease}

\documentclass[twocolumn,amsmath,amssymb,longbibliography,superscriptaddress,aps,pra,10pt]{revtex4-2}
\usepackage{amsmath}
\usepackage[dvipsnames]{xcolor}
\usepackage{multirow}
\usepackage[separate-uncertainty = true,multi-part-units=single]{siunitx}
\usepackage{graphicx}
\usepackage{subfig}
\usepackage[english]{babel}
\usepackage[utf8]{inputenc} 
\usepackage{ctable} 
\usepackage{booktabs}
\usepackage{float}
\usepackage{wasysym}
\usepackage{color, colortbl}
\usepackage{german}
\usepackage[colorlinks=true,allcolors=blue]{hyperref}
\usepackage{doipubmed}
\usepackage{threeparttable}

\newcommand{\Ar}[1]{$^{#1}$Ar}
\newcommand{\Kr}[1]{$^{#1}$Kr}

\definecolor{Lightblue}{rgb}{0.85,0.85,0.95}
\definecolor{Lightred}{rgb}{0.95,0.9,0.9}

\begin{document}
	
	\title{Optically enhanced discharge excitation and trapping of \Ar{39} }
	\author{Y.-Q. Chu}
	\author{Z.-F. Wan}
	\author{F. Ritterbusch}
	\email{florian@ustc.edu.cn}
     \author{W.-K. Hu}
	\author{J.-Q. Gu}
	\author{S.-M. Hu}
	\affiliation{
	University of Science and Technology of China, 96 Jinzhai Road, Hefei 230026, China
	} 
	\author{Z.-H. Jia}
\affiliation{Institute of Modern Physics, Chinese Academy of Sciences, Lanzhou 730000, China}
\affiliation{School of Nuclear Science and Technology, Lanzhou University, Lanzhou 730000, China}
	\author{W. Jiang}
		\affiliation{
	University of Science and Technology of China, 96 Jinzhai Road, Hefei 230026, China
	}

	\author{Z.-T. Lu}
		\email{ztlu@ustc.edu.cn}
		\affiliation{
	University of Science and Technology of China, 96 Jinzhai Road, Hefei 230026, China
	}

	\author{L.-T. Sun}
\affiliation{Institute of Modern Physics, Chinese Academy of Sciences, Lanzhou 730000, China}
\affiliation{School of Nuclear Science and Technology, University of Chinese Academy of Sciences, Beijing 100049, China}
	\author{A.-M. Tong}
			\affiliation{
	University of Science and Technology of China, 96 Jinzhai Road, Hefei 230026, China
	}
     \author{J. S. Wang}
	\author{G.-M. Yang}
	\affiliation{
	University of Science and Technology of China, 96 Jinzhai Road, Hefei 230026, China
	}
	\begin{abstract}
		We report on a two-fold increase of the \Ar{39} loading rate in an atom trap by enhancing the generation of metastable atoms in a discharge source. Additional atoms in the metastable $1s_5$ level (Paschen notation) are obtained via optically pumping both the $1s_4-2p_6$ transition at \SI{801}{nm} and the $1s_2-2p_6$ transition at \SI{923}{nm}. By solving the master equation for the corresponding six-level system, we identify these two transitions to be the most suitable ones and encounter a transfer process between $1s_2$ and $1s_4$ when pumping both transitions simultaneously. We calculate the previously unknown frequency shifts of the two transitions in \Ar{39} and confirm the results with trap loading measurements. The demonstrated increase in the loading rate enables a corresponding decrease in the required sample size, uncertainty and measurement time for \Ar{39} dating, a significant improvement for applications such as dating of ocean water and alpine ice cores.	
	\end{abstract}
	
	
	\maketitle

	\section{Introduction}
\begin{figure*}
\centering
\noindent \includegraphics[width=17cm]{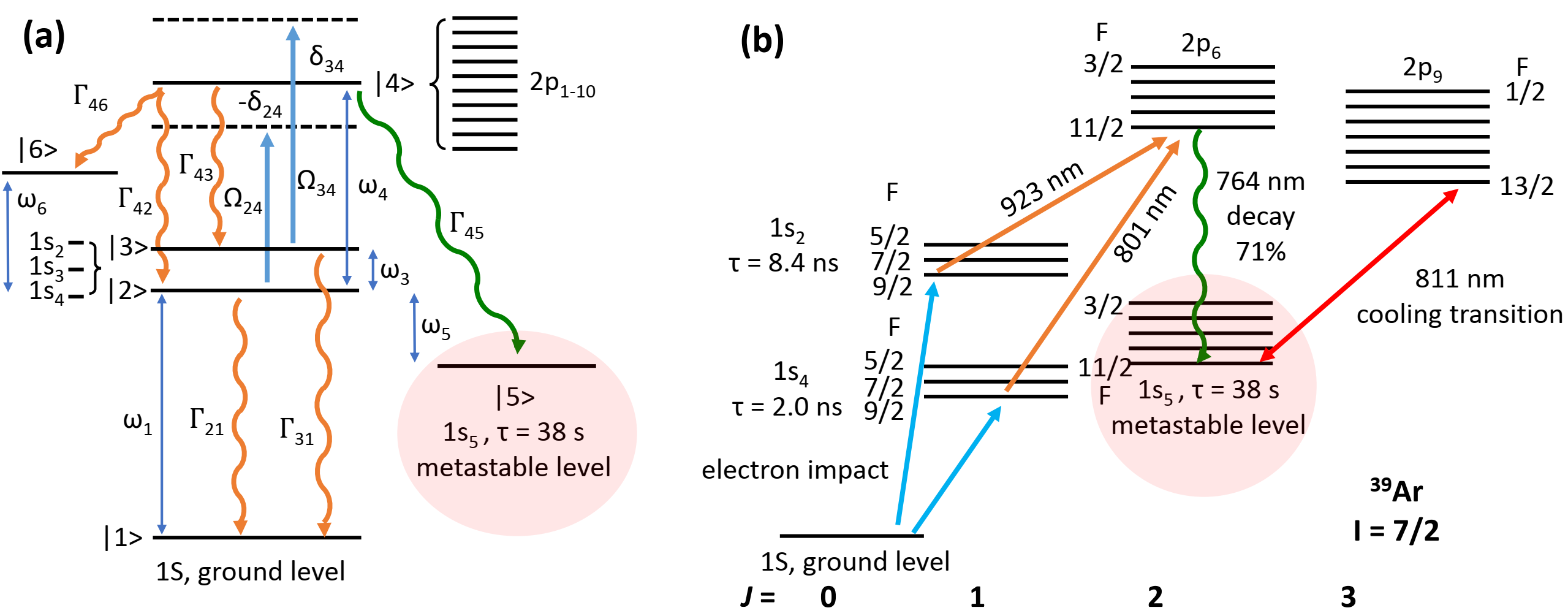}
\caption{
(a) Scheme for enhancing the population in the metastable level $1s_{5}$ by driving the $1s-2p$ levels with Rabi frequencies $\Omega_{24}$, $\Omega_{34}$ and detunings $\delta_{24}$, $\delta_{34}$. $\Gamma_{ij}$ denotes the spontaneous emission rate from level $i$ to $j$. (b) The optical pumping scheme chosen in this work on the $1s_4-2p_6$ transition at \SI{801}{nm} and $1s_2-2p_6$ transition at \SI{923}{nm}, shown for \Ar{39} which has a nuclear spin $I = 7/2$. 
}
\label{fig:scheme}
\end{figure*}

	The noble gas radioisotope \Ar{39} with a half-life of \SI{268\pm8}{years} \cite{Stoenner1965,Chen2018} has long been identified as an ideal dating isotope for water and ice in the age range 50-\SI{1800}{years} due to its chemical inertness and uniform distribution in the atmosphere \cite{Lal1963, Loosli1968}. However, its extremely low isotopic abundances of $10^{-17}-10^{-15}$ in the environment have posed a major challenge in the analysis of \Ar{39}. In the past, it could only be measured by Low-Level Counting, which requires several tons of water or ice \cite{Loosli1983}.\\ \indent In recent years, the sample size for \Ar{39} dating has been drastically reduced by the emerging method Atom Trap Trace Analysis (ATTA), which detects individual atoms via their fluorescence in a magneto-optical trap (MOT). This laser-based technique was originally developed for \Kr{81} and \Kr{85}
 \cite{Chen1999, Jiang2012, Lu2014, Tian2019} and has later been adapted to \Ar{39}, realizing dating of groundwater, ocean water and glacier ice
\cite{Jiang2011,Ritterbusch2014,Ebser2018, Feng2019}. The latest state-of-the-art system reaches an \Ar{39} loading rate of $\sim$\SI{10}{atoms/h} for modern samples and an \Ar{39} background of $\sim$\SI{0.1}{atoms/h} \cite{Tong2021,Gu2021}. Still, its use in applications like ocean circulation studies and dating of alpine glaciers is hampered by the low count rate, which determines the measurement time, precision and sample size.\\ 
\indent Laser cooling and trapping of argon atoms in the ground level is not feasible due to the lack of suitable lasers at the required vacuum ultra violet (VUV) wavelength. As it is the case for all noble gas elements, argon atoms need to be excited to the metastable level $1s_5$ where the $1s_5-2p_9$ cycling transition at \SI{811}{nm} can be employed for laser cooling and trapping (Paschen notation \cite{Paschen1919} is used here, the corresponding levels in Racah notation \cite{Racah1942} can be found in Fig. \ref{transition} in Appendix \ref{app:ar_transitions}). The $1s_5$ level is $\sim$\SI{10}{eV} above the ground level and, in operational ATTA instruments, is populated by electron-impact excitation in a RF-driven discharge with an efficiency of only $ 10^{-4}-10^{-3} $. Increasing this efficiency would raise the loading rate of \Ar{39} accordingly. \\	
\indent Since the discharge excites atoms into not only the metastable $1s_5$ but also many other excited levels, the metastable $1s_5$ population can be enhanced by transferring atoms from these other excited levels to the metastable $1s_5$ via optical pumping (Fig. \ref{fig:scheme}). This mechanism has been demonstrated in a spectroscopy cell for argon with an increase of 81\% \cite{Hans2014,Frolian2015} and for xenon with an increase by a factor of 11 \cite{Hickman2016, Lamsal2020}. It has also been observed in an argon beam with an increase of 21\% \cite{Hans2014}. While these experiments were done on stable and abundant isotopes, a \SI{60}{\%} increase in loading rate has recently been observed for the rare isotopes \Kr{81} and \Kr{85} \cite{Zhang2020}.\\
\indent In this work, we theoretically and experimentally examine the enhancement of metastable production by optical pumping for the rare \Ar{39} as well as the abundant argon isotopes. We identify the $1s_4-2p_6$ transition at \SI{801}{nm} and the $1s_2-2p_6$ transition at \SI{923}{nm} as the most suitable candidates. Implementing the enhancement scheme for \Ar{39} on these transitions requires knowing the respective frequency shifts, which we calculate and experimentally confirm. Moreover, loading rate measurements support the theoretically predicted transfer process between $1s_2$ and $1s_4$ levels when driving the 923-nm and 801-nm transitions simultaneously.
	
\subsection{Transfer efficiency}\label{theory_transfer_efficiency}
We solve the Lindblad master equation (see details in Appendix \ref{Lindblad_master_equations}) for the 6-level system shown in Fig. \ref{fig:scheme}(a) which corresponds to the even argon isotopes without hyperfine structure. The resulting steady-state solution $\widetilde{\rho}_{55}(t\to+\infty)$ for the final population in the metastable level
can be obtained analytically as a function of the initial populations in $ {\left |2 \right\rangle} $ and $ {\left |3 \right\rangle} $, using the initial condition
\begin{equation}
	\widetilde{\rho}_{ij}(t=0)=0\text{\ \ for \ }(i,j)\not=(2,2) \ \text{and} \  (i,j)\not=(3,3).
\end{equation}

If only one transition is driven, e.g. $\Omega_{34}=0$, then $\widetilde{\rho}_{55}(t\to+\infty)$ simplifies to the expressions given in \cite{Zhang2020}. We use these expressions to calculate the transfer efficiency $ \widetilde{\rho}_{55}(t\to+\infty) $ for the different $ 1s-2p $ transitions in even argon isotopes as a function of laser power. The transitions with the highest transfer efficiencies are shown in Table \ref{tab:transfer_efficiencies} (see Table \ref{tab:trans_frac} in Appendix \ref{argon_transition_scheme_trans_frac} for all transitions). 
\begin{table}[H]
\caption{
Argon transitions with the highest transfer efficiencies from each $ 1s $ level, calculated for a laser beam of 9-mm diameter and different powers $P$. For driving both transitions simultaneously (bottom row) an equal population in $1s_2$ and $1s_4$ and equal power $P$ for each laser beam is assumed.
}
\centering
\def\arraystretch{1.3} 
\begin{tabular*}{\hsize}{@{}@{\extracolsep{\fill}}ccccc@{}}
\hline\hline
\multirow{2}{*}{\begin{tabular}[c]{@{}c@{}}Lower\\ level\end{tabular}} & \multirow{2}{*}{\begin{tabular}[c]{@{}c@{}}Upper\\ level\end{tabular}} & \multirow{2}{*}{$ \lambda \SI{}{(nm)}$} & \multicolumn{2}{c}{$ \widetilde{\rho}_{55}(t\to+\infty) $} \\ \cline{4-5} 
			&                              &                     & $ P=\SI{0.5}{W} $      & $ P\rightarrow +\infty \SI{}{W} $          \\ \hline
			$ 1s_{4} $                          & $ 2p_{6} $                          & 801              & 0.03       & 0.05      \\ 
			$ 1s_{3} $                          & $ 2p_{10} $                        & 1047            & 0.77       & 0.77      \\ 
			$ 1s_{2} $                          & $ 2p_{6} $                          & 923              & 0.15       & 0.17      \\ \hline
			$ 1s_{2}+1s_{4} $           & $ 2p_{6} $                  & 801+923     & 0.12       &0.08        \\ \hline
\hline
\end{tabular*}
\label{tab:transfer_efficiencies}
\end{table}	
	From the metastable level $1s_3$ (see Fig. \ref{transition} in Appendix \ref{argon_transition_scheme_trans_frac}), the $1s_3-2p_{10}$ transition at \SI{1047}{nm} has the highest transfer efficiency of \SI{77}{\%}. Since $1s_3$ is also metastable, only a few mW of laser power are needed to saturate the transition. However, experimentally we only achieve an increase in the metastable $1s_5$ population of $\sim$\SI{10}{\%} by pumping this transition. Since the increase in the population of the metastable $1s_5$ is the product of the transfer efficiency (=0.77, Table \ref{tab:transfer_efficiencies}) and the initial population in the $1s_3$, it follows that the latter is only $10\%/0.77=13\% $ of that in the metastable $1s_5$. Given this limitation, optical pumping on $1s_3$ is not investigated further in this work.\\
\indent The transfer efficiency from $1s_2$ is the highest for the 923-nm transition to $2p_{6}$, reaching a high-power limit of \SI{17}{\%}. From $1s_4$ the transfer efficiency is the highest for the 801-nm transition to $2p_{6}$, reaching a high-power limit of \SI{5}{\%}. Since the populations of these levels in the argon discharge are not known, the actual increase in the metastable population needs to be determined experimentally. In the following we focus on these two transitions as illustrated in Fig. \ref{fig:scheme}(b) for the odd isotope \Ar{39}.\\  
 \begin{figure}[h!]
\centering
\vspace{0.3cm}
\noindent \includegraphics[width=0.4\textwidth]{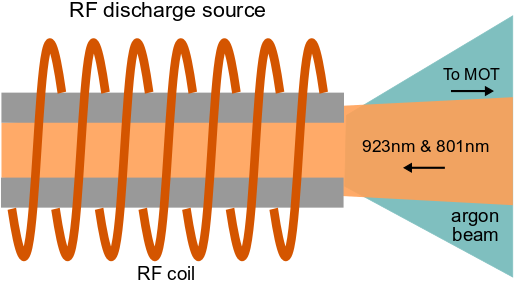}
\caption{RF-driven discharge source of metastable argon atoms in the ATTA setup. The optical pumping laser beams are sent into the source counter-propagating to the atomic beam. }
\label{fig:setup}
\end{figure}

 \begin{figure*}[t!]
  \centering
  \subfloat{
 \label{801freq.1}
 \includegraphics[width=0.5\textwidth]{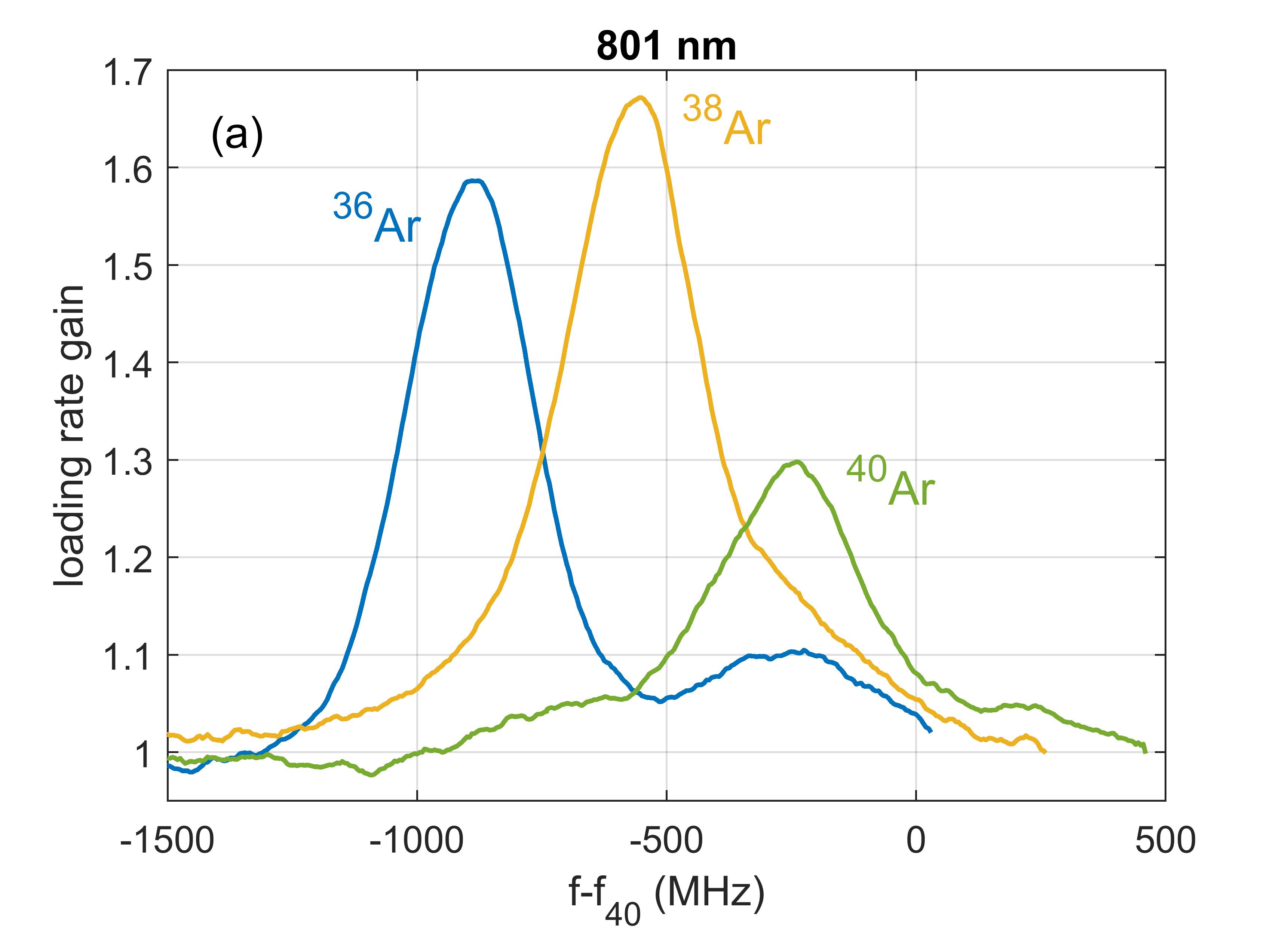}}
 \subfloat{
 \label{922freq.2}
 \includegraphics[width=0.5\textwidth]{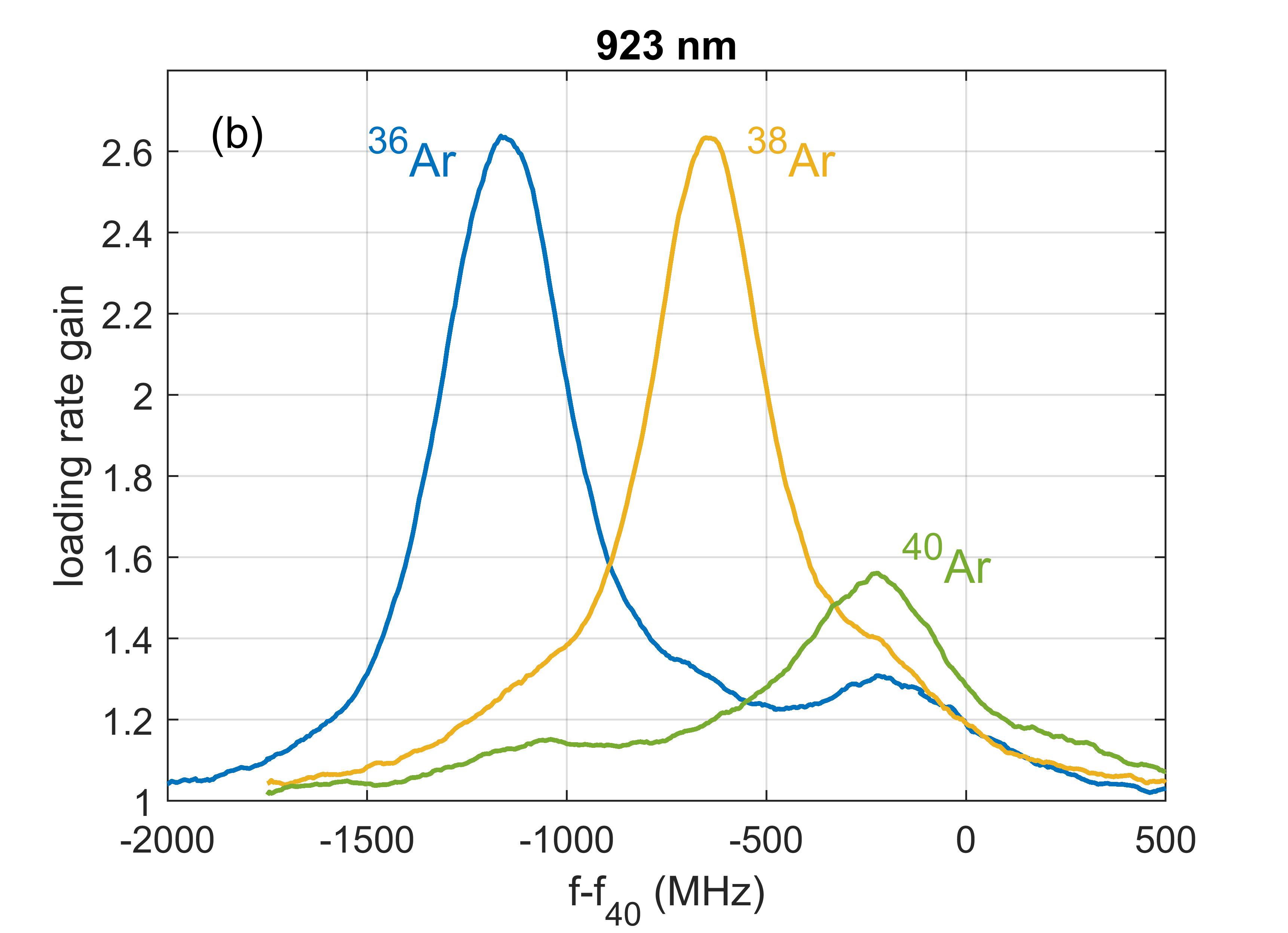}}
 \captionsetup{justification=raggedright}
 \caption{Gain in the MOT loading rate for the abundant argon isotopes vs frequency of the (a) 801-nm and (b) 923-nm optical pumping light, measured in the enriched sample. For each transition, $f_{40}$ denotes the resonance frequency of \Ar{40} at rest as monitored in a spectroscopy cell.}
 \label{fig:even isotopes}
 \end{figure*}
 
   \indent Interestingly, when these two transitions are driven simultaneously (i.e. $\Omega_{24}\neq0 , \Omega_{34}\neq0$) the final population in the metastable level $\rho_{55}$ is smaller than the sum of the individually driven transitions (see bottom row of table \ref{tab:transfer_efficiencies}). This effect is the consequence of stimulated emission from $2p_6$ to $1s_4$ by the 801-nm light, together with the 923-nm light effectively transferring atoms from $1s_2$ to $1s_4$. In the same way atoms are also transferred from $1s_4$ to $1s_2$. However, since the decay rate to the ground level from $1s_4$ is three times higher than from $1s_2$ (see Fig. \ref{transition} in Appendix \ref{app:ar_transitions}), the total increase in the metastable level is lower than the sum of the individually driven transitions. As the laser power increases also the stimulated emission increases, leading to a further decrease in the combined transfer efficiency to the metastable level.
 	
\subsection{Isotope shifts and hyperfine splittings for \Ar{39}} \label{iso_hyper}
The total frequency shifts of \Ar{39} for the 801-nm and 923-nm transitions consist of the isotope shifts and the hyperfine splittings. The hyperfine coefficients of \Ar{39} for $1s_2$ and $1s_4$ were measured in \cite{Traub1967}, whereas for $2p_6$ they can be calculated from the corresponding hyperfine coefficients measured for \Ar{37} \cite{Klein1996}, using the measured nuclear magnetic dipole moments and electric quadrupole moments of \Ar{39} and \Ar{37}
 \cite{Armstrong1971,Zhang2020,Stone2015,Klein1996}. The resulting hyperfine coefficients are shown in Table \ref{tab:A_B} in Appendix \ref{A_B_ar_cal}. Isotope shifts of neither the 801-nm transition nor the 923-nm transition for any argon isotope have been found in the literature. 
The isotope shifts for \Ar{36} and \Ar{38} have therefore been measured in this work (see below), allowing us to calculate the isotope shifts for \Ar{39} \cite{King1963, Heilig1974, Zhang2020}. The resulting isotope shifts and hyperfine splittings for \Ar{39} relative to \Ar{40} are given in Table \ref{tab:ar_cal} in Appendix \ref{A_B_ar_cal}.

\section{Experimental Setup}
For measuring the metastable population increase by optical pumping in \Ar{39} as well as the stable argon isotopes, we use an ATTA system as described in \cite{Tong2021}.
Metastable argon atoms are generated in a RF-driven discharge by electron impact (Fig. \ref{fig:setup}) and are subsequently laser cooled and detected in a magneto-optical trap. 
Single \Ar{39} atoms are detected via their 811-nm fluorescence in the MOT using an electron-multiplying charged coupled device (EMCCD) camera. During a measurement of \Ar{39} (\Ar{39}/Ar=$8\times 10^{-16}$ in modern air), the stable and abundant \Ar{38} (\Ar{38}/Ar=\SI{0.06}{\%} in air) is measured as well to account for drifts in the trap loading efficiency. The loading rate of \Ar{38} for this normalization purpose is measured by depopulating the MOT with a quenching transition and detecting the emitted fluorescence \cite{Jiang2012, Tong2021}. For testing optical pumping on \Ar{38} and the other stable argon isotopes the loading rate is measured by first clearing the MOT with a quenching transition and then the initial linear part of the rising slope of the MOT fluorescence is measured \cite{Cheng2013}. \\
\indent For optical pumping, we shine in the \SI{923}{nm} and \SI{801}{nm} laser beams counter-propagating to the atomic beam (Fig. \ref{fig:setup}). The laser beams are weakly focused and slightly larger than the inner diameter of the source tube (\diameter \SI{10}{mm}). 
\begin{figure*}[t!]
  \subfloat{
 \label{801hyperfine.1}
\hspace{-0.6cm}
 \includegraphics[width=0.55\textwidth]{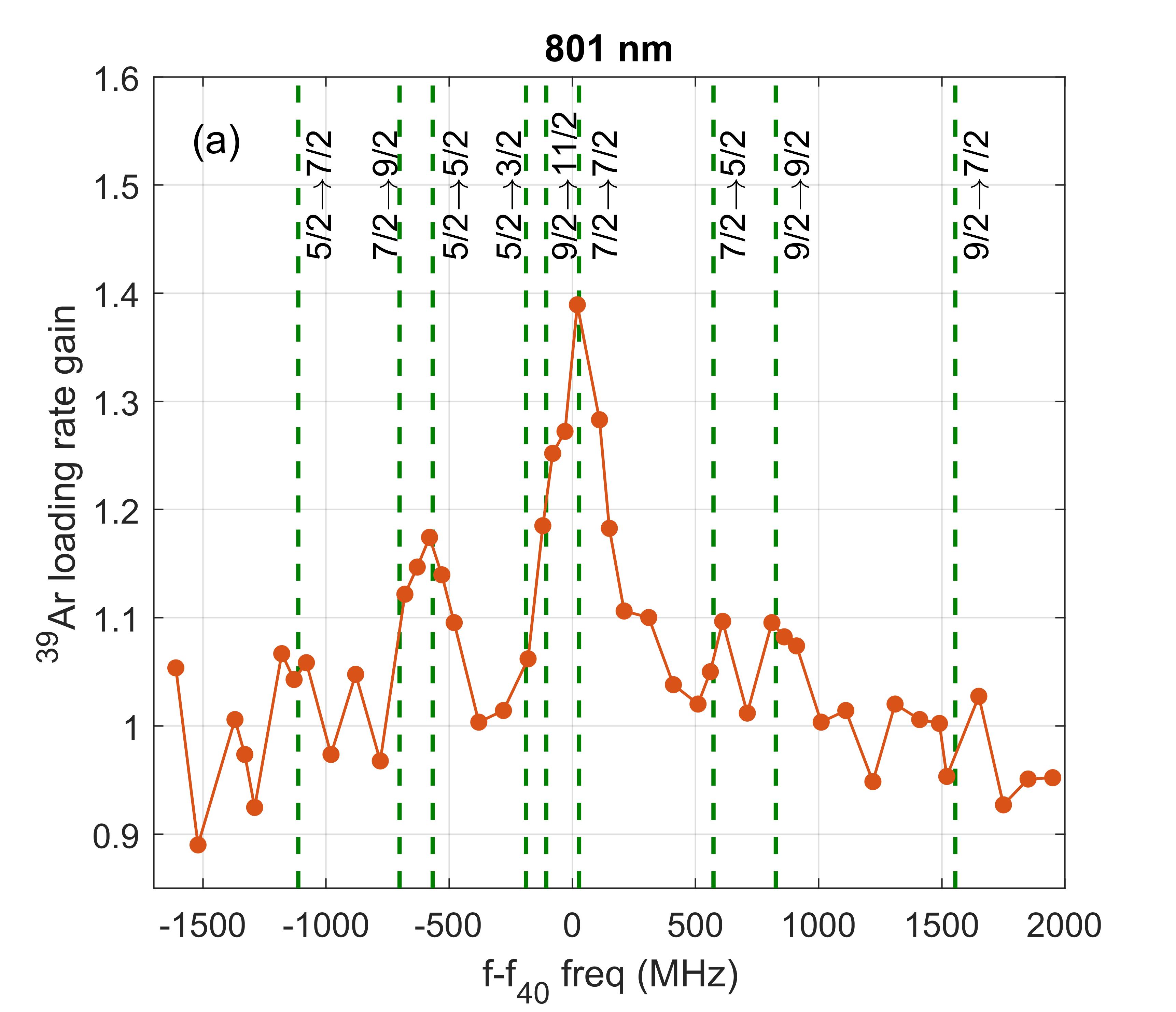}}
 \subfloat{
 \label{923hyperfine.2}
\hspace{-0.8cm}
 \includegraphics[width=0.55\textwidth]{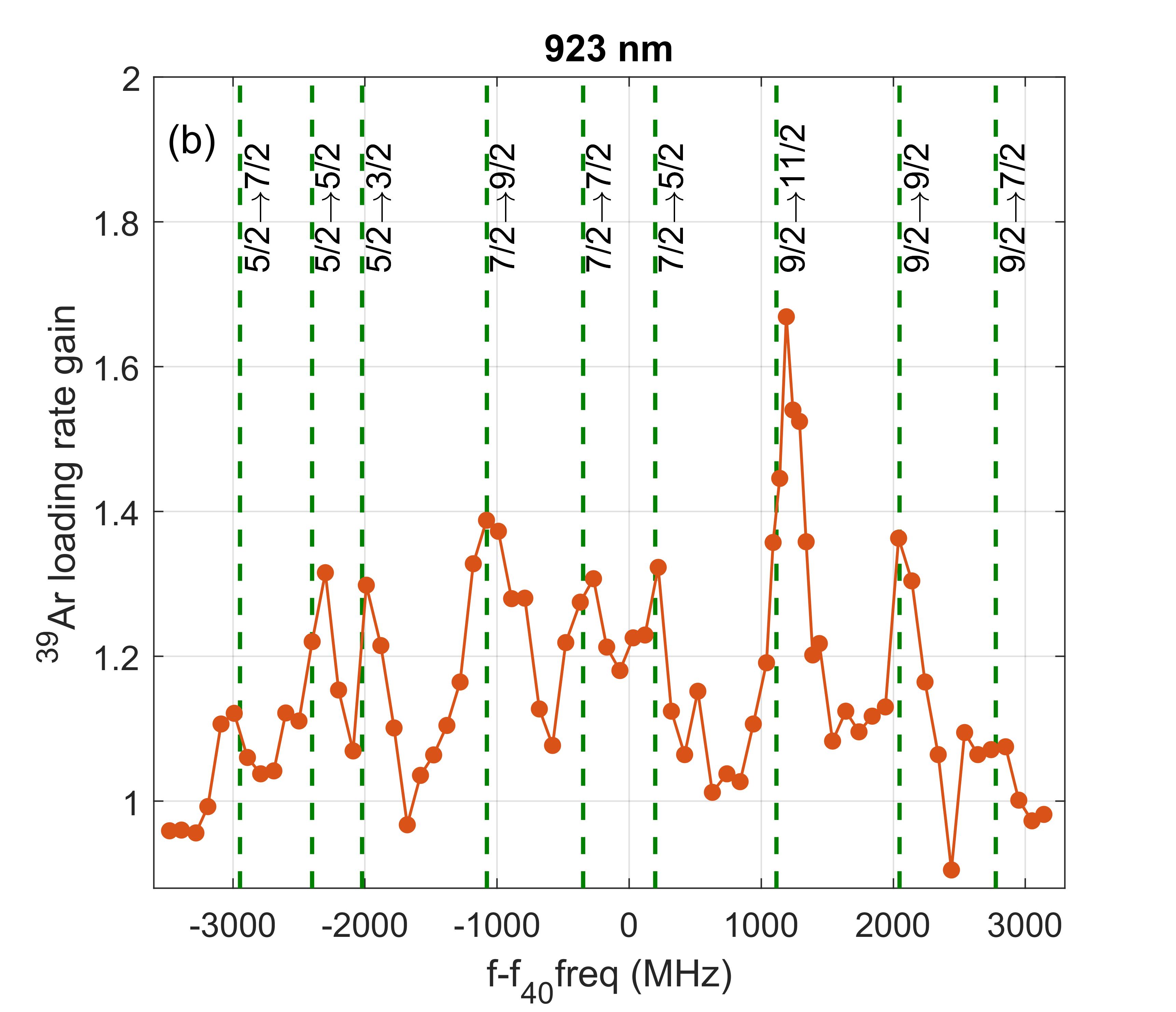}}
 \captionsetup{justification=raggedright}
 \caption{\Ar{39} loading rate gain versus frequency of the (a) 801-nm and (b) 923-nm light, measured in the enriched sample. $f_{40} $ denotes the resonance frequency of \Ar{40} at rest as monitored in a spectroscopy cell. The Doppler-shift obtained for \Ar{40} has been subtracted from the frequency to obtain the \Ar{39} frequency spectrum at rest. The error of each \Ar{39} data point is $\sim$\SI{5}{\%}. The dashed green lines indicate the calculated frequencies of the hyperfine transitions.}
 \label{fighyperfine}
 \end{figure*}
The optical pumping light is generated by tapered amplifiers seeded with diode lasers, providing up to \SI{1.0}{W} of usable laser power at \SI{801}{nm} and \SI{1.6}{W} at \SI{923}{nm}. For measuring the different argon isotopes, the laser frequency needs to be tuned and stablilized over several GHz. For this purpose, the two lasers are locked by a scanning transfer cavity lock \cite{Zhao1998, Subhankar2019}, using a diode laser locked to the 811-nm cooling transition of metastable \Ar{40} as the master. In order to increase counting statistics for \Ar{39}, we use an enriched sample prepared by an electromagnetic mass separation system \cite{Jia2020}. In the enriched sample, \Ar{40} is largely and \Ar{36} partially removed so that \Ar{39} and \Ar{38} are enriched by a factor $\sim200$. The ratio of \Ar{39} and \Ar{38} is not changed in the enrichment process \cite{Tong2022}, which is important for the normalization described above.
The \Ar{40}, \Ar{36} and \Ar{38} abundances in the enriched sample are \SI{60}{\%}, \SI{30}{\%} and \SI{10}{\%}, respectively. 
	
\section{Results and Discussion}	
\begin{figure}[h!]
\centering
\noindent \includegraphics[width=0.5\textwidth]{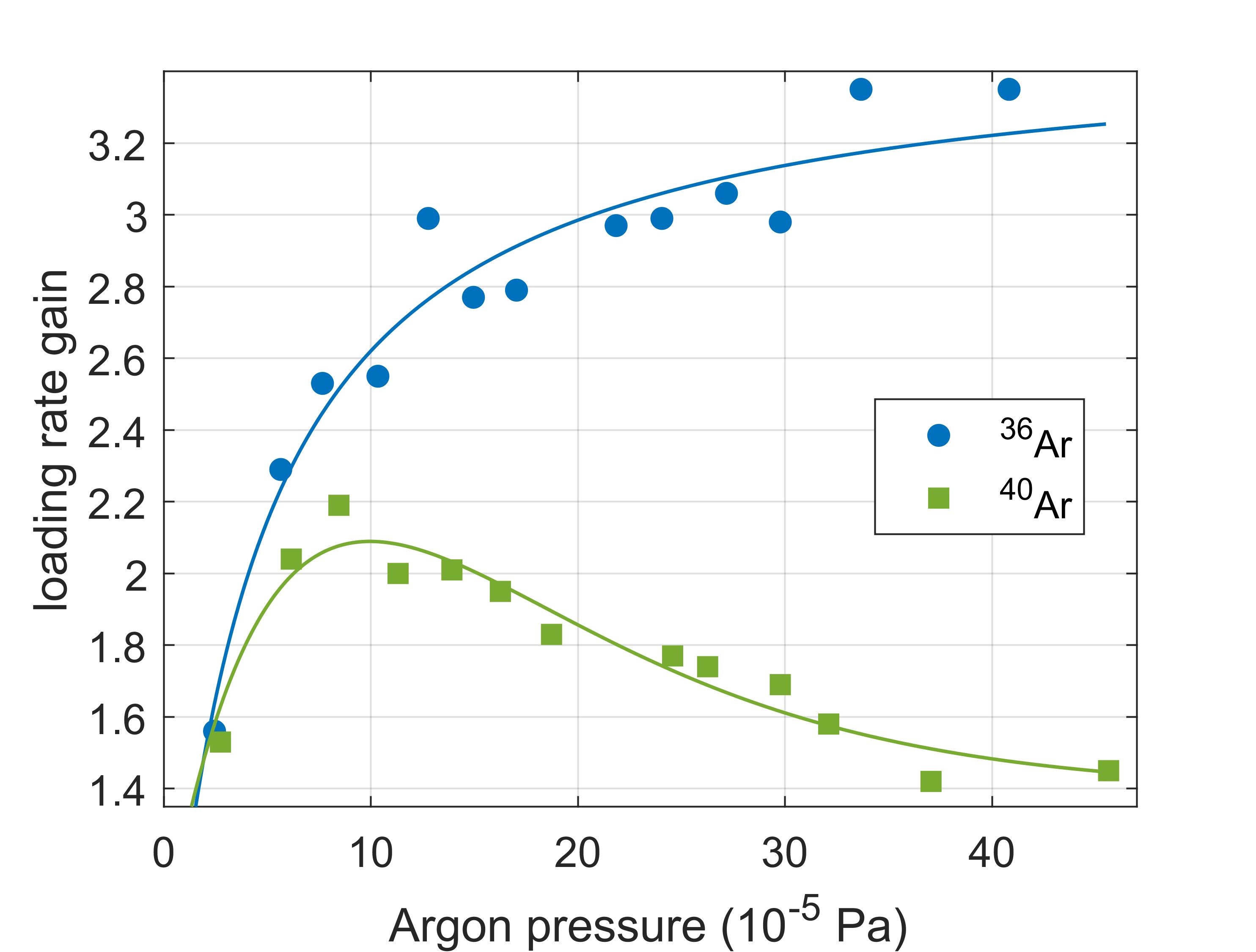}
\caption{Loading rate gain vs argon pressure in the chamber at the outlet of the source tube (Fig. \ref{fig:setup}) for the 923-nm transition, measured with atmospheric argon. The pressure inside the source tube is considerably higher than at the outlet of the source tube. The lines are guides-to-the-eye. }
\label{fig:gain_vs_pressure}
\end{figure}

The loading rate of the stable argon isotopes is measured versus the frequencies of the 801-nm and 923-nm light (Fig. \ref{fig:even isotopes}). A clear increase in the loading rate is observed for all isotopes. For \Ar{40} we obtain most probable Doppler shifts around \SI{-230}{MHz} in agreement with the expected temperature of the liquid-nitrogen-cooled atomic beam. The small \Ar{40} feature mirrored on the positive detuning side is likely caused by the optical pumping light reflected at the window behind the source. The window is partially coated by metal which has been sputtered by argon ions that are produced in the discharge.
From the observed resonances for \Ar{36} and \Ar{38} we obtain the isotope shifts with respect to \Ar{40} for the 801-nm as well as the 923-nm transition shown in Table \ref{tab:isotope shifts}. Based on these measured isotope shifts, we calculate the isotope shifts for \Ar{39} (Table \ref{tab:isotope shifts}) using King plots \cite{King1963}. Interestingly, the loading rate of \Ar{36} shows a pronounced increase also at the \Ar{40} resonance for both, the 801-nm and the 923-nm transition. Looking closely, an increase in loading rate is visible for each isotope at the resonances of the other two isotopes. This additional increase is likely caused by metastable exchange collisions, e.g. transferring an increase in the metastable population of \Ar{40} to that of \Ar{36}. The maximum loading rate gain is lower for \Ar{40} than for the less abundant \Ar{36} and \Ar{38}. This difference is discussed in more detail below.\\
\begin{figure*}[t!]
  \centering
  \subfloat{
 \label{801power.1}
\hspace{-0.8cm}
 \includegraphics[width=0.56\textwidth]{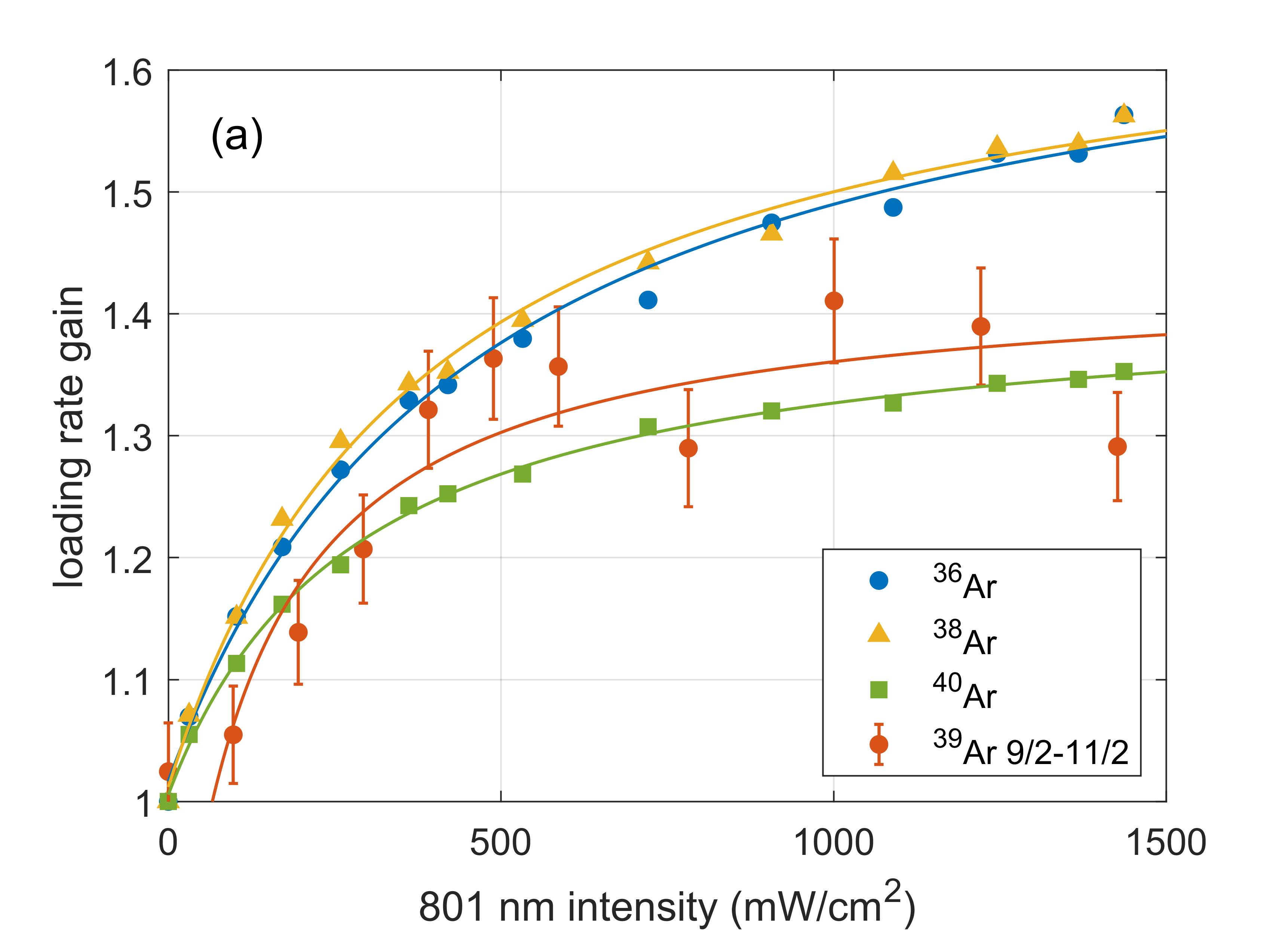}}
\subfloat{
 \label{922power.2}
\hspace{-0.8cm}
 \includegraphics[width=0.56\textwidth]{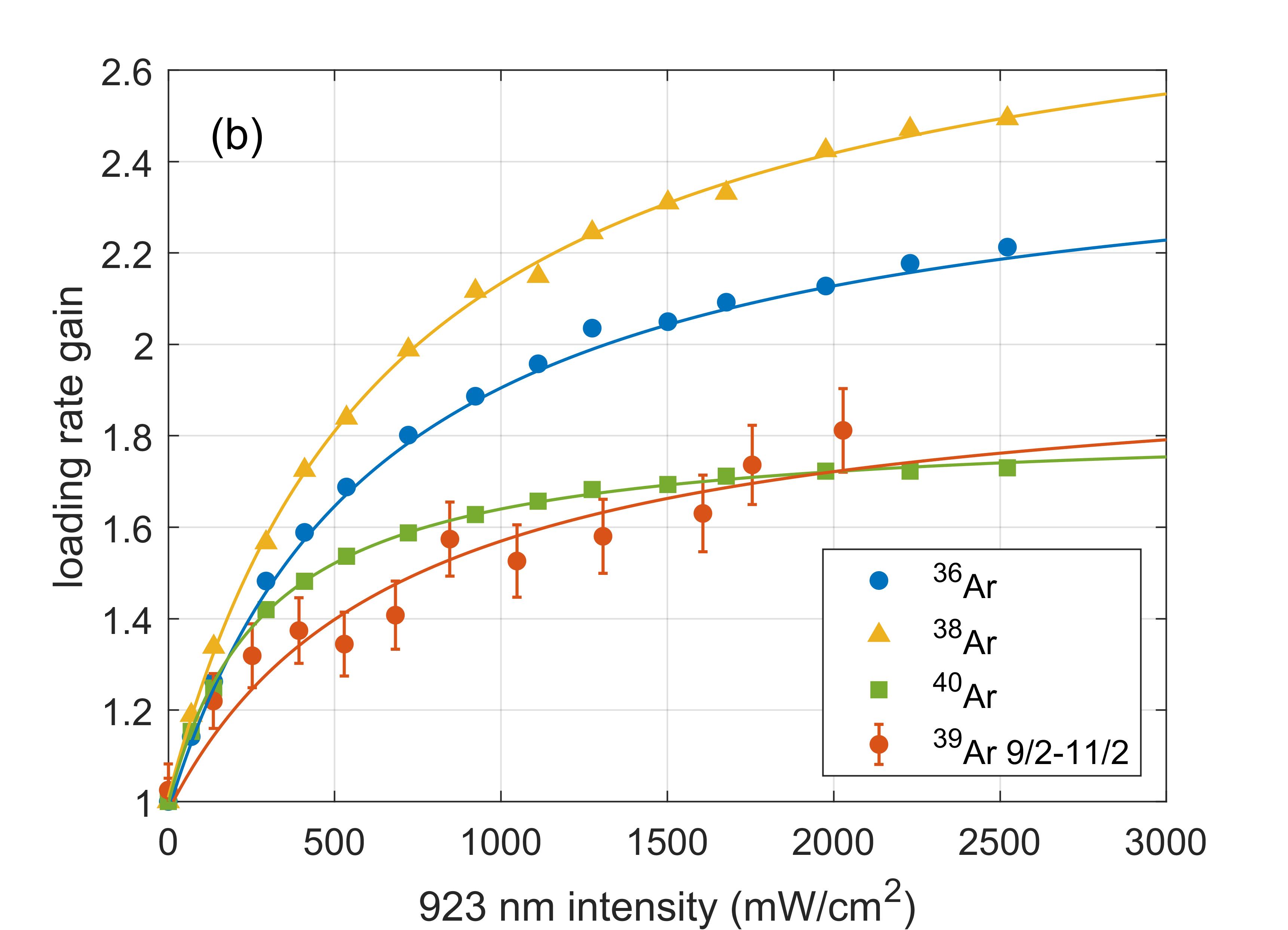}}
 \captionsetup{justification=raggedright}
 \caption{(a) Loading rate gain of the argon isotopes vs laser power of the (a) 801-nm light and (b) 923-nm light, measured with the enriched argon sample. The lines are saturation fits according to the expressions given in \cite{Zhang2020}. }
 \label{fig.power}
\end{figure*}
\indent Fig. \ref{fighyperfine} shows the \Ar{39} loading rate gain vs frequency of the 801-nm and 923-nm light. For both transitions, a clear increase in the loading rate is observed. For \SI{923}{nm}, the $F=9/2\rightarrow11/2$ transition is the strongest as expected from the multiplicity and transition strength \cite{Axner2004}. Moreover, the measurements confirm the other calculated hyperfine transitions. For \SI{801}{nm}, the overlap of the $F=9/2\rightarrow11/2 $ and $ F=7/2\rightarrow7/2 $ transition is the strongest. The loading rate increase is lower compared to that achieved with the 923-nm light. Accordingly, the different hyperfine transitions are resolved less clearly. Nevertheless, the measurements are in good agreement with the calculated hyperfine transitions. In order to address not only one but two hyperfine levels of \Ar{39}, we add sidebands to the 801-nm and 923-nm light. At \SI{801}{nm} no increase is detectable by adding a sideband resonant with the overlap of the $F=7/2\rightarrow9/2 $ and $ F=5/2\rightarrow5/2 $ transitions. At \SI{923}{nm} we observe a maximum increase of only $\sim$\SI{10}{\%}, although according to Fig. \ref{923hyperfine.2} an increase of \SI{40}{\%} appears possible. Likely, the increase by adding a sideband is compensated by the decrease due to the lower laser power on the carrier frequency. \\
\indent The loading rate gain as a function of laser power is shown in Fig. \ref{fig.power}. As already observed in Fig. \ref{801freq.1}, the maximum loading rate gain is lower for \Ar{40} than for the less abundant \Ar{36} and \Ar{38}. This may be caused by the higher density of \Ar{40} leading to a stronger trapping of the \SI{764}{nm} fluorescence (see Fig. 
 \ref{fig:scheme}), which can quench other metastable atoms. Moreover, the saturation intensity is significantly lower for \Ar{40} than for \Ar{36} and \Ar{38}. This may also be caused by the higher density of \Ar{40}, leading to trapping of the re-emitted 801-nm and 923-nm light. The saturation intensity for \Ar{39} is difficult to assess due to the large measurement uncertainties and the contribution from neighbouring hyperfine levels. \\
\indent Table \ref{tab:loading_rate} lists the maximum loading rate gains of the different argon isotopes for the 801-nm and the 923-nm transitions, as well as for both transitions driven together. As predicted by the calculation in section \ref{theory_transfer_efficiency} and Appendix \ref{Lindblad_master_equations}, driving
\begin{table}[h]
\caption{Loading rate gains obtained for different argon isotopes and different transitions, measured in the enriched argon sample.}
\centering
\def\arraystretch{1.3} 
\begin{tabular*}{\hsize}{@{}@{\extracolsep{\fill}}cccc@{}}
\hline\hline
Isotope                                                         & \SI{801}{nm}    & \SI{923}{nm}    & \SI{801}{nm} + \SI{923}{nm} \\ \hline
\begin{tabular}[c]{@{}l@{}}\Ar{40} \end{tabular} &1.4                  &1.7      & \begin{tabular}[c]{@{}l@{}}1.8\end{tabular} \\ 
\begin{tabular}[c]{@{}l@{}}\Ar{38} \end{tabular} & 1.6                 & 2.5     & 2.8                                            \\   
\begin{tabular}[c]{@{}l@{}}\Ar{36} \end{tabular} & 1.6                 & 2.2      & 2.6                                           \\ 
\Ar{39}                                            & 1.4         & 1.8   & 2.0 \\ \hline\hline
\end{tabular*}
\label{tab:loading_rate}
\end{table}
 both transitions simultaneously results in a lower gain than the addition of the individual gains. This result confirms the transfer due to stimulated emission between $1s_2$ and $1s_4$ via the intermediate $2p_6$, driven by the 923-nm and 801-nm light. For \Ar{39} a two-fold gain in the loading rate is obtained by optical pumping when simultaneously using 801-nm and 923-nm light and addressing the $F=9/2$ level. According to the loading rate gain obtained for the other hyperfine levels (Fig. \ref{fighyperfine}), if sidebands are introduced with additional laser power  a near three-fold gain in the loading rate should be possible. \\
\indent As mentioned above and observed in Fig. \ref{fig.power}, the loading rate gain varies for different isotopes. Moreover, we observe that the loading rate gain depends on density and sample composition. In order to examine the dependence, we measure the \Ar{36} and \Ar{40} loading rate gains vs. argon pressure in the chamber at the outlet of the source tube (Fig. \ref{fig:gain_vs_pressure}). In this measurement, atmospheric argon (abundances of \Ar{40}, \Ar{36} and \Ar{38} are \SI{99.6}{\%}, \SI{0.33}{\%} and \SI{0.06}{\%}, respectively) is used instead of the enriched sample. 
The loading rate gains of the two isotopes differ significantly. For \Ar{36} the loading rate gain increases with the argon pressure whereas for \Ar{40} the loading rate gain decreases beyond a maximum. Moreover, the loading rate gain for \Ar{36} in this measurement with atmospheric argon reaches the value 3.3 whereas it is only 2.2 when measured with the enriched sample (\Ar{36} abundance=\SI{30}{\%}) as in Fig. \ref{fig.power}. These findings indicate that the populations of the $1s$-levels in the discharge depend on pressure and composition. These dependences might be caused by various mechanisms such as trapping of light from the VUV ground level transitions, which together with the optical pumping light can produce metastable argon atoms. 
	
\section{Conclusion and Outlook}
We have realized a two-fold increase of the \Ar{39} loading rate in an atom trap system via optical pumping in the discharge source. A three-fold increase is expected by adding sidebands with additional laser power that cover all the hyperfine levels of \Ar{39}. Similarly, we obtain an increase of the MOT loading rate by a factor 2-3 for the stable argon isotopes \Ar{36}, \Ar{38} and \Ar{40}.
We observe that the loading rate gain varies for different isotopes and that it depends on the argon pressure in the discharge as well as the abundance of the respective isotope. We assign these dependences to the complex population dynamics of the $1s$-levels in the discharge via mechanisms such as radiation trapping and metastable exchange collisions. Consequently, using the method presented here for practical \Ar{39} analysis requires a stable control of the pressure so that the loading rate gain due to optical pumping for both \Ar{39} and \Ar{38} stays constant during measurements. \\
\indent The hitherto unknown isotope shifts in \Ar{36} and \Ar{38} as well as the \Ar{39} spectra for the 801-nm and 923-nm transitions have been measured in this work. They constitute an important contribution to the efforts on optically generating metastable argon via resonant two-photon excitation \cite{Wang2021, Dong2022}. For a more precise measurement of the hyperfine coefficients and the isotope shifts, spectroscopy on samples highly enriched in \Ar{39} will be necessary \cite{Welte2009, Williams2011}. \\
\indent The presented method for enhanced production of metastable argon can be directly implemented in existing ATTA setups to increase the \Ar{39} loading rate by a factor 2-3. For state-of-the-art ATTA systems, the \Ar{39} loading rate is $\sim$\SI{10}{atoms/h}. For \Ar{39} analysis at a precision level of \SI{5}{\%}, this loading rate leads to a measuring time of $\sim$\SI{50}{h} during which the \Ar{39} background in the ATTA system increases linearly with time. Therefore, the two-fold increase in \Ar{39} loading rate realized in this work constitutes a significant advance for measuring time, precision and sample size of \Ar{39} analysis in environmental applications such as dating of alpine ice cores and large scale ocean surveys.

%

	\begin{acknowledgments}This work is funded by the National Natural Science Foundation of China (41727901, 41961144027, 41861224007), National Key Research and Development Program of China (2016YFA0302200), Anhui Initiative in Quantum Information Technologies (AHY110000).  \\
\\
\indent Y.-Q. Chu and Z.-F. Wan contributed equally to this work.\\
\\
\indent \textit{ An edited version of this paper was published by APS \href{https://journals.aps.org/pra/abstract/10.1103/PhysRevA.105.063108}{Physical Review A 105, 063108 (2022)}. Copyright 2022 American Physical Society.}
	\end{acknowledgments}

%
	
\clearpage
	
\appendix

\onecolumngrid

\section{Argon $1s - 2p$ transitions}	\label{app:ar_transitions}

\begin{figure}[h]
\centering \vspace{-0.5cm}
\noindent \includegraphics[width=20cm,angle=270]{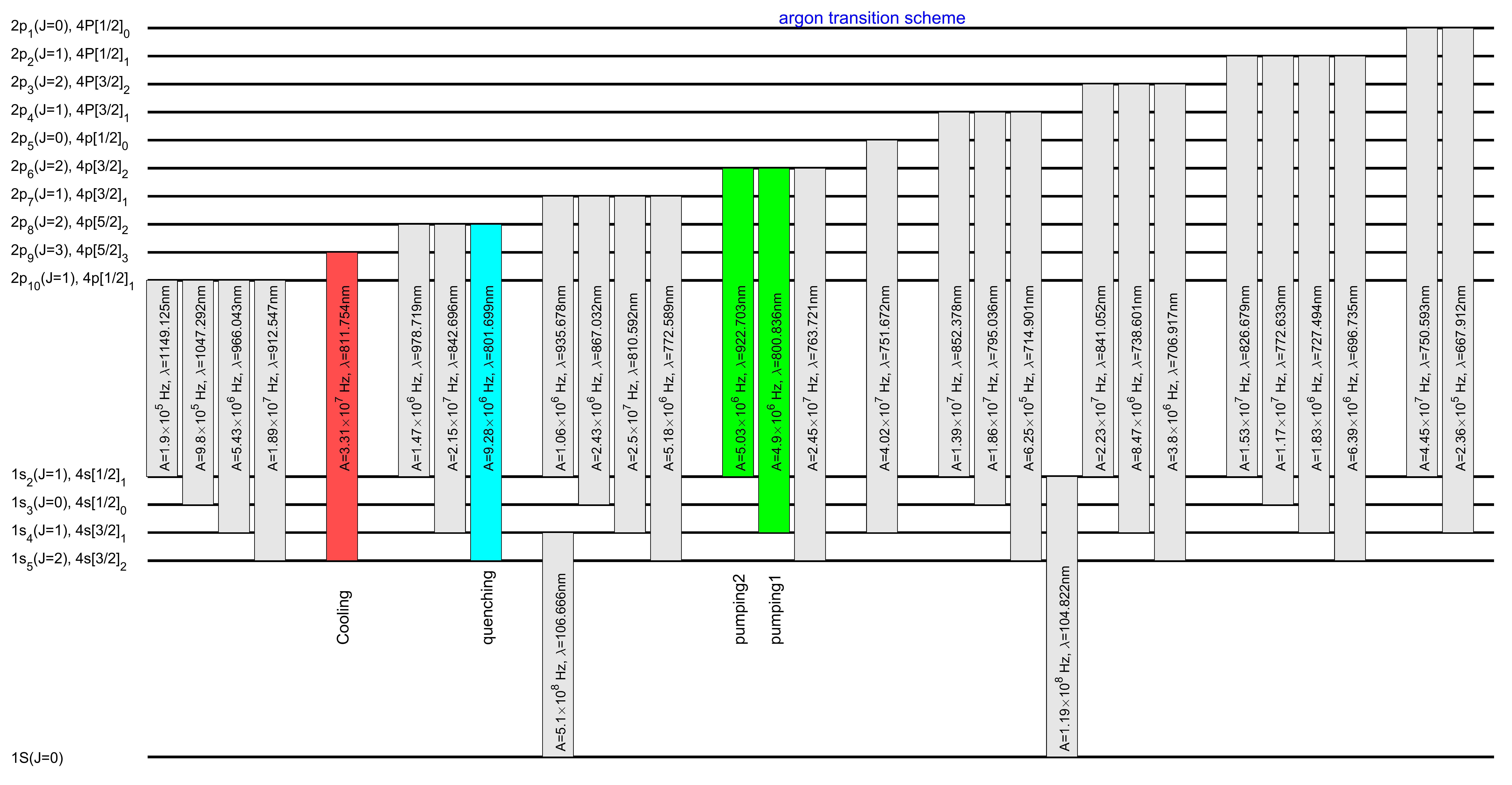}
\caption{
Argon transition scheme calculated based on \cite{NIST_ASD} and adopted from \cite{ritterbusch2009, welte2011}.
 Levels with capital letter in the Racah notation refer to $ j_{\text{core}}=1/2 $ while levels with small letters refer to $ j_{\text{core}}=3/2 $.
}
\label{transition}
\end{figure}

\clearpage
\twocolumngrid

	\section{Master equation}\label{Lindblad_master_equations}
	The 6-level system for optical pumping of the even argon isotopes without hyperfine structures is illustrated in Fig. \ref{fig:scheme}. As described in section \ref{theory_transfer_efficiency}, $\vert1\rangle$ is the ground level and $\vert5\rangle$ the metastable level for laser cooling and trapping. Atoms in levels $\vert2\rangle$ and $\vert3\rangle$ can be transferred to $\vert5\rangle$ by driving the transition to $\vert4\rangle$ followed by spontaneous decay. $\vert6\rangle$ represents other levels that atoms can decay to from $\vert4\rangle$.
Choosing the energy of level $\vert2\rangle$ as zero, $-\hbar\omega_1, \hbar\omega_3, \hbar\omega_4, -\hbar\omega_5$ and $\hbar\omega_6$ are the energies of the corresponding levels relative to $\vert2\rangle$.
	In the interaction picture, the Hamiltonian of this atomic system interacting with the laser field is
	
	\begin{equation}
	\hat{H}=\hat{H}_A+\hat{H}_{AF},
	\end{equation} 
	where 
	\begin{eqnarray}
      \begin{split}
	\hat{H}_A &= \hbar\omega_4\vert 4\rangle\langle 4\vert-\hbar\omega_1\vert 1\rangle\langle 1\vert+\hbar\omega_3\vert 3\rangle\langle 3\vert \\
	&-\hbar\omega_5\vert 5\rangle\langle 5\vert+\hbar\omega_6\vert 6\rangle\langle 6\vert
              \end{split}
	\end{eqnarray}
	is the atomic Hamiltonian and 
	\begin{eqnarray}
           \begin{split}
	\hat{H}_{AF} &=&\frac{\hbar}2(\Omega_{24}^*\sigma_{24}e^{i\omega_{24} t}+\Omega_{24}\sigma_{42}e^{-i\omega_{24} t}) \\
	& & + \frac{\hbar}2(\Omega_{34}^*\sigma_{34}e^{i\omega_{34} t}+\Omega_{34}\sigma_{43}e^{-i\omega_{34} t}) 
           \end{split}
	\end{eqnarray} 
	is the Hamiltonian that describes the interaction of the atoms with the light field. Here, $\omega_{24}$ and $\omega_{34}$ are the laser frequencies of the incident light, $\Omega_{24}$ and $\Omega_{34}$ are the corresponding Rabi frequencies and $\sigma_{ij}=\vert i\rangle\langle j\vert$ are the spin operators. 
	With the unitary transformation $U=\mathrm{exp}(i\omega_{24} t\vert 4\rangle\langle 4\vert + i(\omega_{24} - \omega_{34}) t \vert 3\rangle\langle 3\vert)$, the quantum level $\vert\psi\rangle$ changes to
	\begin{equation}
	\widetilde{\vert\psi\rangle}=U\vert\psi\rangle.
	\end{equation}
	
	In this Schr{\"o}dinger picture, the Hamiltonian becomes
	\begin{equation}
	\begin{split}
	\widetilde{H}&=UHU^\dagger+i\hbar(\partial_tU)U^\dagger\\
	&=-\hbar\delta_{24}\sigma_{44}-\hbar\omega_1\sigma_{11}-\hbar\omega_5\sigma_{55}+\hbar\omega_6\sigma_{66} \\
	&+\hbar(\delta_{34}-\delta_{24})\sigma_{33}+\frac{\hbar}2(\Omega_{24}\sigma_{42}+\Omega_{34}\sigma_{43}+h.c.)  \\
	\end{split}
	\end{equation}
	where $\delta_{24}=\omega_{24}-\omega_4$ and $\delta_{34}=\omega_{34}-(\omega_4-\omega_3)$ are the detunings of the light with respect to the transition frequencies.
	\par 
	The Lindblad master equation for the system including the spontaneous emission can be written as
	\begin{equation} 
\begin{split}
\frac{d\widetilde{\rho}}{dt}
&=\frac{1}{i\hbar}[\widetilde{H}, \widetilde{\rho}]\\
&\quad+\Gamma_{42}(\sigma_{24}\widetilde{\rho}\sigma_{42}-\frac12\widetilde{\rho}\sigma_{44}-\frac12\sigma_{44}\widetilde{\rho})\\
&\quad+\Gamma_{21}(\sigma_{12}\widetilde{\rho}\sigma_{21}-\frac12\widetilde{\rho}\sigma_{22}-\frac12\sigma_{22}\widetilde{\rho})\\
&\quad+\Gamma_{45}(\sigma_{54}\widetilde{\rho}\sigma_{45}-\frac12\widetilde{\rho}\sigma_{44}-\frac12\sigma_{44}\widetilde{\rho})\\
&\quad+\Gamma_{46}(\sigma_{64}\widetilde{\rho}\sigma_{46}-\frac12\widetilde{\rho}\sigma_{44}-\frac12\sigma_{44}\widetilde{\rho})\\
&\quad+\Gamma_{43}(\sigma_{34}\widetilde{\rho}\sigma_{43}-\frac12\widetilde{\rho}\sigma_{44}-\frac12\sigma_{44}\widetilde{\rho})\\
&\quad+\Gamma_{31}(\sigma_{13}\widetilde{\rho}\sigma_{31}-\frac12\widetilde{\rho}\sigma_{33}-\frac12\sigma_{33}\widetilde{\rho})\\
\end{split}
\end{equation}
	where $\Gamma_{ij}$ is the spontaneous emission rate from $\vert i\rangle$ to $\vert j\rangle$. These equations describe the time evolution of $\widetilde{\rho}_{ij}=\langle i\vert\widetilde{\rho}\vert j\rangle$ and can be simplified to
\begin{align}\label{eq:time_evolution}
\frac{d\widetilde{\rho}_{11}}{dt}&=\Gamma_{21}\widetilde{\rho}_{22} +\Gamma_{31}\widetilde{\rho}_{33}  \nonumber  \\
\frac{d\widetilde{\rho}_{22}}{dt}&=\Gamma_{42}\widetilde{\rho}_{44}-\Gamma_{21}\widetilde{\rho}_{22}+\frac{i}{2}(\Omega_{24}\widetilde{\rho}_{24}-\Omega^{*}_{24}\widetilde{\rho}_{42})  \nonumber  \\
\frac{d\widetilde{\rho}_{44}}{dt}&=-(\Gamma_{42}+\Gamma_{45}+\Gamma_{46}+\Gamma_{43})\widetilde{\rho}_{44}  \nonumber  \\
&+\frac{i}{2}(\Omega^{*}_{24}\widetilde{\rho}_{42}-\Omega_{24}\widetilde{\rho}_{24}) +\frac{i}{2}(\Omega^{*}_{34}\widetilde{\rho}_{43}-\Omega_{34}\widetilde{\rho}_{34})  \nonumber   \nonumber  \\
\frac{d\widetilde{\rho}_{55}}{dt}&=\Gamma_{45}\widetilde{\rho}_{44}  \nonumber  \\
\frac{d\widetilde{\rho}_{66}}{dt}&=\Gamma_{46}\widetilde{\rho}_{44}   \\
\frac{d\widetilde{\rho}_{33}}{dt}&=\Gamma_{43}\widetilde{\rho}_{44} -\Gamma_{31} \widetilde{\rho}_{33}+\frac{i}{2}(\Omega_{34}\widetilde{\rho}_{34}-\Omega^{*}_{34}\widetilde{\rho}_{43})   \nonumber  \\
\frac{d\widetilde{\rho}_{42}}{dt}&=-\frac12(\Gamma_{42}+\Gamma_{21}+\Gamma_{45}+\Gamma_{46}+\Gamma_{43}-2i\delta_{24})\widetilde{\rho}_{42} \nonumber \\
& +\frac{i\Omega_{24}}{2} (\widetilde{\rho}_{44}-\widetilde{\rho}_{22})-\frac{i\Omega_{34}}{2}\widetilde{\rho}_{32}    \nonumber  \\
\frac{d\widetilde{\rho}_{43}}{dt}&=-\frac12(\Gamma_{42}+\Gamma_{31}+\Gamma_{45}+\Gamma_{46}+\Gamma_{43}-2i\delta_{34})\widetilde{\rho}_{43}  \nonumber\\
&+ \frac{i\Omega_{34}}{2} (\widetilde{\rho}_{44} -\widetilde{\rho}_{33})-\frac{i\Omega_{24}}{2}\widetilde{\rho}_{23}  \nonumber  \\
\frac{d\widetilde{\rho}_{32}}{dt}&=-\frac12(\Gamma_{21}+\Gamma_{31}+2i(\delta_{34}-\delta_{24}))\widetilde{\rho}_{32}  \nonumber  \\
&-\frac{i\Omega^{*}_{34}}{2}\widetilde{\rho}_{42}+\frac{i\Omega_{24}}{2}\widetilde{\rho}_{34} \nonumber
\end{align}
	Using that the population is initially in $\vert2\rangle$ and $\vert3\rangle$, i.e. 
	\begin{equation}
	\begin{split}
	\widetilde{\rho}_{22}(t=0) &\neq 0  \ , \  \widetilde{\rho}_{33}(t=0) \neq 0   \\
	\widetilde{\rho}_{ij}(t=0)&=0 , \text{ }(i,j)\not=(2,2), (i,j)\not=(3,3), 
	\end{split}
	\end{equation}
	then for the steady-state
	\begin{equation}
	\frac{d\widetilde{\rho}}{dt}(t\rightarrow+\infty)=0
	\end{equation}
	Eq. \ref{eq:time_evolution} can be solved analytically using a computer algebra system, yielding the transfer efficiency $\widetilde{\rho}_{55}(t\to+\infty)$.
	
	\clearpage
	
	\onecolumngrid
	
		\section{TRANSFER EFFICIENCIES FOR $ 1s–2p $ TRANSITIONS IN \Ar{40}}\label{argon_transition_scheme_trans_frac}
		
		To determine the most suitable transitions for optical pumping to the metastable level $ 1s_5 $, we have theoretically investigated all the $ 1s-2p $ transitions in argon. The transfer efficiency for each transition has been calculated according to the derivation in Sec. \ref{theory_transfer_efficiency} and the results are compiled in Table \ref{tab:trans_frac}. For each $ 1s $ level we can thereby identify the transition with the highest transfer efficiency. Among these, we experimentally find the $ 1s_{2}-2p_{6} $ transition at \SI{923}{nm} and the $1s_{4}-2p_{6} $ transition at \SI{801}{nm} to be the strongest ones for optical pumping and therefore have chosen them for this work. A scheme of all $ 1s-2p $ transitions in argon is illustrated in Fig. \ref{transition} with the levels in Racah as well as in Paschen notation.
		\begin{table}[H]
			\centering	
			\def\arraystretch{1.3} 
			\caption{Transfer efficiencies $ \widetilde{\rho}_{55}(t\to+\infty) $ for $ 1s-2p $ transitions in \Ar{40} calculated for a laser beam with 9-mm diameter and different powers $ P $. The transitions highlighted in bold are the ones with the highest transfer efficiency. The levels are provided in Paschen as well as in Racah notation.}
			\begin{tabular*}{0.618\hsize}{@{}@{\extracolsep{\fill}}ccccc@{}}
				\\ \hline\hline
				\multirow{2}*{Lower level}&\multirow{2}*{Upper level}&\multirow{2}*{$\lambda \ \SI{}{(nm)}$}&\multicolumn{2}{c}{Transfer efficiency $ \widetilde{\rho}_{55}(t\to+\infty) $} \\ \cline{4-5} &&&$P=\SI{0.5}{W}$&$ P\rightarrow +\infty \SI{}{W} $  \\ \hline
				\multirow{7}*{$1s_4$, $4s[3/2]_1$}&$2p_{10}$, $4p[1/2]_1$&966.04&0.03&0.04 \\
				&$2p_8$, $4p[5/2]_2$&842.70&0.02&0.02\\ 
				&$2p_7$, $4p[3/2]_1$&810.60&0.01&0.01\\ 
				&\boldmath{$2p_6$, $4p[3/2]_2$}&\textbf{800.84}&\textbf{0.03}&\textbf{0.05}  \\ 
				&$2p_4$, $4P[3/2]_1$&747.12&0.00&0.00\\ 
				&$2p_3$, $4P[3/2]_2$&738.60&0.00&0.01 \\ 
				&$2p_2$, $4P[1/2]_1$&727.50&0.00&0.01
				\\ \hline
				\multirow{4}*{$1s_3$, $4s[1/2]_0$}
				&\boldmath{$2p_{10}$, $4p[1/2]_1$}&\textbf{1047.30}&\textbf{0.77}&\textbf{0.77} \\ 
				&$2p_7$, $4p[3/2]_1$&867.03&0.17&0.17\\ 
				&$2p_4$, $4P[3/2]_1$&795.04&0.04&0.04\\ 
				&$2p_3$, $4P[1/2]_1$&772.63&0.27&0.27
				\\ \hline
				\multirow{7}*{$1s_2$, $4s[1/2]_1$}&$2p_{10}$, $4p[1/2]_1$&1149.13&0.05&0.13\\
				&$2p_8$, $4p[5/2]_2$&978.72&0.05&0.06 \\ 
				&$2p_7$, $4p[3/2]_1$&935.68&0.02&0.03 \\

				&\boldmath{$2p_6$, $4p[3/2]_2$}&\textbf{922.70}&\textbf{0.15}&\textbf{0.17} \\ 
				&$2p_4$, $4P[3/2]_1$&852.38&0.00&0.00\\ 
				&$2p_3$, $4P[3/2]_2$&841.05&0.03&0.03\\ 
				&$2p_2$, $4P[1/2]_1$&826.68&0.04&0.05\\  \hline\hline
			\end{tabular*}
			\label{tab:trans_frac}
		\end{table}

\clearpage

	\section{ISOTOPE, HYPERFINE, AND TOTAL FREQUENCY SHIFTS FOR THE 801-nm AND 923-nm TRANSITION IN \Ar{39}}\label{A_B_ar_cal}	
	Realizing optical pumping for the odd argon isotopes requires knowledge of the frequency shifts for the employed 801-nm and 923-nm transition. The total frequency shift is the sum of the isotope shift and the hyperfine shift. The isotope shifts for \Ar{39} have not been measured and were calculated based on the measured isotope shifts for the stable isotopes (see Sec. \ref{iso_hyper}). The resulting isotope shifts for \Ar{36}, \Ar{38} and \Ar{39} are shown in Table \ref{tab:isotope shifts}.	The hyperfine constants of the involved levels have been measured for \Ar{39} or can be calculated from measurements for \Ar{37} (see Sec. \ref{iso_hyper}). The resulting hyperfine shifts for the different hyperfine levels are compiled in Table \ref{tab:ar_cal} together with the isotope shifts and the total frequency shifts. 

\begin{table}[h]
\centering
\def\arraystretch{1.3}
\caption{Isotope shifts relative to \Ar{40} for the 801-nm and 923-nm transitions.}
 \begin{threeparttable}
\begin{tabular*}{0.45\hsize}{@{}@{\extracolsep{\fill}}cccccc@{}}  
			\\ \hline\hline
			\begin{tabular}[c]{@{}c@{}}Transition\\ nm\end{tabular}        & \begin{tabular}[c]{@{}c@{}}Isotope\end{tabular}   & \begin{tabular}[c]{@{}c@{}}Isotope shift\\ MHz\end{tabular}  \\ \hline
			\multirow{3}{*}{801}    & \Ar{36}      & $ -640(10)^{\text a} $     \\ 
			                        & \Ar{38}      & $ -290(10)^{\text a} $     \\ 
                                        & \Ar{39}      & $ -128(16)^{\text b} $     \\ \hline
                 \multirow{3}{*}{923}   & \Ar{36}      & $ -960(10)^{\text a} $      \\ 
			                        & \Ar{38}      & $ -445(10)^{\text a} $      \\ 
                                        & \Ar{39}      & $ -202(16)^{\text b} $      \\   \hline\hline                      
\end{tabular*}
  \begin{tablenotes}
  \footnotesize
    \item[a] Measured in this work.
     \item[b]Calculated in this work.
  \end{tablenotes}
\end{threeparttable}
\label{tab:isotope shifts}
\end{table}


\begin{table}[H]
		\centering
		\def\arraystretch{1.3}
		\caption{Hyperfine coefficients $ A $ and $ B $ for \Ar{39} and different levels.}
          \begin{threeparttable}
		\begin{tabular*}{0.5\hsize}{@{}@{\extracolsep{\fill}}cccc@{}}  
			\hline\hline
			\begin{tabular}[c]{@{}c@{}}Wavelength\\ nm\end{tabular}                                & level          & \begin{tabular}[c]{@{}c@{}}$A$\\MHz\end{tabular} & \begin{tabular}[c]{@{}c@{}}B\\MHz\end{tabular} \\ \hline
			\multirow{2}{*}{801} & $ 1s_4 $ & $ -334(2)^{\text a} $ & $ -24(3)^{\text a} $ \\  
			& $ 2p_6 $ & $ -163(31)^{\text b} $ & $ -63(11)^{\text b} $   \\ \hline
			\multirow{2}{*}{923} & $ 1s_2 $ & $ -712(1)^{\text a} $ & $ 84(6)^{\text a} $ \\  
			& $ 2p_6 $ & $ -163(31)^{\text b} $ & $  -63(11)^{\text b} $    \\  \hline\hline
		\end{tabular*}
   \begin{tablenotes}
     \footnotesize
      \item[a]Reference \cite{Traub1967}.
      \item[b]Reference \cite{Klein1996}.
      \end{tablenotes}
    \end{threeparttable}
\label{tab:A_B}
\end{table}


	\begin{table}[h]
		\centering
		\def\arraystretch{1.3}
		\caption{Isotope, hyperfine and total frequency shifts for the 801-nm and 923-nm transitions in \Ar{39}. The hyperfine shift is relative to the center of gravity of the fine-structure term and the isotope shift is relative to \Ar{40}.}
		\begin{tabular*}{0.8\hsize}{@{}@{\extracolsep{\fill}}cccccc@{}}  
			\\ \hline\hline
			\begin{tabular}[c]{@{}c@{}}Transition\\ nm\end{tabular} & \begin{tabular}[c]{@{}c@{}}Isotope shift\\ MHz\end{tabular}           & \begin{tabular}[c]{@{}c@{}}Lower level\\ $ 1s_4 $/$ 1s_2 $\end{tabular}              & \begin{tabular}[c]{@{}c@{}}\hspace{0.5cm}Upper level\\ $ 2p_6 $\end{tabular}  & \begin{tabular}[c]{@{}c@{}}HFS shift\\ MHz\end{tabular} & \begin{tabular}[c]{@{}c@{}}Total shift\\ MHz\end{tabular} \\ \hline
			\multirow{9}{*}{801} & \multirow{9}{*}{$ -128(16) $} & \multirow{3}{*}{$ F=5/2 $}  & $ F=3/2 $       & -61(275)   &-189(275)    \\ 
			&                         &                         & $ F=5/2 $       & -440(198)    & $  -567(199) $       \\ 
			&                         &                         & $ F=7/2 $       & $ -985(92) $    & $  -1113(93)$        \\ \cline{3-6}
			&                         & \multirow{3}{*}{$ F=7/2 $}  & $ F=5/2 $       &701(198)     & 573 (199)       \\ 
			&                         &                         & $ F=7/2 $       & 155(92)      & $ 27(93) $         \\ 
			&                         &                         & $ F=9/2 $       & $ -574(46)$    & $  -702(49)$      \\ \cline{3-6}
			&                         & \multirow{3}{*}{$ F=9/2 $}  & $ F=7/2 $       & 1682(92)    & 1555(93)      \\ 
			&                         &                         & $ F=9/2 $       & 953(46)   & 826(49)      \\ 
			&                         &                         & $ F=11/2 $      & $ 21(214)$    & $ -106(214)$    \\ \cline{1-6} 
			\multirow{9}{*}{923} & \multirow{9}{*}{$ -202(16) $}  & \multirow{3}{*}{$ F=5/2 $}  & $ F=3/2 $       &  -1820(275)     & -2022(275)       \\ 
			&                         &                         & $ F=5/2 $       & $ -2198(198) $       & $-2401(199)  $       \\ 
			&                         &                         & $ F=7/2 $       & $   -2744(92)$    & $-2946(93)$      \\ \cline{3-6}
			&                         & \multirow{3}{*}{$ F=7/2 $}  & $ F=5/2 $       & 400(198)     & 197(199)       \\ 
			&                         &                         & $ F=7/2 $       &  -146(92)     &  -348(93)      \\ 
			&                         &                         & $ F=9/2 $      & $ -875(46) $      & $ -1077(49) $      \\ \cline{3-6}
			&                         & \multirow{3}{*}{$ F=9/2 $} & $ F=7/2 $       &  2978(92)      & 2776(93)      \\ 
			&                         &                         & $ F=9/2 $      & 2249(46)    & 2047(49)        \\ 
			&                         &                         & $ F=11/2 $      & $ 1317(214) $    & $1115(214)$       \\ \hline\hline                          
		\end{tabular*}
		\label{tab:ar_cal}
	\end{table}

	\newpage
	\twocolumngrid
	\bibliographystyle{apsrev4-2}
	\bibliography{ref}


\end{document}